# Run-Time-Reconfigurable Multi-Precision Floating-Point Matrix Multiplier Intellectual Property Core on FPGA

S. Arish[1] · R. K. Sharma[1]

**Abstract** In today's world, high-power computing applications such as image processing, digital signal processing, graphics, robotics require enormous computing power. These applications use matrix operations, especially matrix multiplication. Multiplication operations require a lot of computational time and are also complex in design. We can use field-programmable gate arrays as low-cost hardware accelerators along with a low-cost general-purpose processor instead of a high-cost application-specific processor for such applications. In this work, we employ an efficient Strassen's algorithm for matrix multiplication and a highly efficient run-time-reconfigurable floating-point multiplier for matrix element multiplication. The run-time-reconfigurable floating-point multiplier is implemented with custom floating-point format for variable-precision applications. A very efficient combination of Karatsuba algorithm and Urdhva Tiryagbhyam algorithm is used to implement the binary multiplier. This design can effectively adjust the power and delay requirements according to different accuracy requirements by reconfiguring itself during run time.

## 1 Introduction

Matrix multiplication is the most significant operation in high-power computing applications. Multiplication operations require a lot of computational time, and the architecture of such multiplication units is complex. The complexity of conventional matrix multiplication algorithm is $O(N^3)$ [1], where $N$ is the order of the matrix. For such processing applications, we cannot use a low-cost general-purpose data processor. An application-specific signal processor is necessary. Also, most matrix multiplication algorithms are implemented with sequential logic in microcontrollers or digital signal processors (DSPs) using sequential languages such as C, C++ or with sequential tools such as MATLAB. This sequential programming can increase the execution time multi-fold. By using FPGAs as low-cost hardware accelerators, we can replace the high-cost application-specific processor with a low-cost general-purpose processor. Also by using a very efficient matrix multiplication algorithm like Strassen's algorithm [6] having a complexity of $O(N^{2.81})$, we can further reduce the complexity of the architecture. Since FPGAs are highly parallel and reconfigurable, we can reconfigure the hardware efficiently according to different applications and the execution speed can be reduced significantly. Even though reconfigurability is the main feature of FPGA, run-time-reconfigurability is an entirely different term. Reprogramming the device again and again while usage is not feasible, and there comes the importance of run-time-reconfigurability. If we can reconfigure the device on the run according to the application requirements, it can be a very efficient solution. Floating-point units can be used instead of fixed point units to increase the flexibility and the range of operation.

✉ S. Arish
arishsu@gmail.com

[1] School of VLSI Design and Embedded Systems, National Institute of Technology Kurukshetra, Kurukshetra, Haryana, India

A lot of effort has been made over the past few decades to improve the performance of floating-point computations. Floating-point units not only are complex but also require more area and hence consume more power as compared to fixed point units. The complexity of the floating-point unit increases when we design for high accuracy. Even a minute error in accuracy can cause major consequences. These errors are possible in floating-point units mainly because of the discrete behaviour of the IEEE-754 [11] floating-point representation, where fixed number of bits is used to represent numbers. Due to the high computational requirements of scientific applications such as computational geometry, climate modelling, computational physics, it is necessary to have extreme precision in floating-point calculations. And this increased precision may not be provided with single-precision or double-precision floating-point format. These factors further increase the complexity of the unit. But some applications do not require high precision. Even an approximate value will be sufficient for the correct operation. For applications which require lower precision, the use of double-precision or quadruple-precision floating-point units will be a luxury. It wastes area, power and also increases latency. For devices such as portable or wearable devices where accuracy requirement varies with different applications and also where power consumption is a crucial factor, the use of high-precision floating-point multipliers is not a good option. In such cases, a variable-precision multiplier can be handy, which can save much power and time when the application does not require high precision. A power efficient design of floating-point multiplier with different modes of accuracy selection is presented and proposed in this paper. With different precision modes, the mode which is most appropriate for the concerned application can be selected. As accuracy or precision requirement decreases, the width of the multiplier decreases and hence the power consumption and latency.

## 2 Literature Review

There are many pieces of literature available on matrix multiplication on the FPGA-based platform and also a few available on floating-point matrix multiplication. The mathematical model for the matrix multiplication algorithm based on Baugh–Wooley algorithm is described in the paper [1]. This model also uses a systolic architecture with parallel processing elements. But as matrix size increases, delay significantly increases. These architectures are useful for multiplication of sub-matrices in Strassen algorithm. An excellent review of different matrix multiplication algorithms including classical matrix multiplication algorithms, Winograd algorithm and Strassen algorithm is explained in detail in the paper [6]. Matrix multiplication described in [5] is for 3D affine transformations, and the main features are the proposed floating-point (FP) multiplier and adder, where it uses a pipelined FP multiply–accumulate unit. A codesign approach for matrix multiplication using conventional algorithm is described in paper [14]. It uses parallel adders and multipliers. Since it uses the conventional algorithm, significant improvement in delay cannot be guaranteed. A floating-point matrix multiplication algorithm using a systolic architecture is explained in paper [21]. A parallel implementation of large matrix multiplier coprocessor is described in paper [4]. A highly parallel matrix product coprocessor is explained in [8], and it is also suitable for large matrix multiplication. But most of these pieces of literature do not really concentrate on variable precision or reconfigurability. The proposed paper focus mainly on reconfigurability.

Different variable-precision floating-point methods are explained in papers [2,7,15]. A multi-mode floating-point multiplier, which operates efficiently with every precision format specified by the IEEE 754-2008 standard, is presented in paper [15]. The proposed multiplier in paper [15] is pipelined to achieve execution of one quadruple multiplication in 3 cycles. A model and implementation of a versatile multiplier, which can perform either double-precision (paired) or single-precision floating-point multiplications, or 16-bit or 8-bit SIMD integer (vector) multiplications, is presented in paper [7]. An efficient variable-precision floating-point multiplier design which requires very less storage area and also uses parallel additions along with multiplication is described in paper [2]. All of these studies illustrate very efficient variable-precision design methods, but run-time-reconfigurability is not the main feature of these models. Readily available IPs such as DSP blocks are also used often.

A combination of Karatsuba algorithm [16] and Urdhva Tiryagbhyam algorithm is used to implement floating-point multiplier in the proposed model. References [11,17] give an idea of floating-point formats. The proposed paper follows the IEEE-754 standard as described in [11]. A very good explanation of various floating-point formats and multiplier algorithms is given in [17]. The implementation of Karatsuba algorithm is described in papers [2,16]. A simple implementation of Karatsuba algorithm is explained in paper [16]. Different Vedic mathematics methods including Urdhva Tiryagbhyam algorithm are described in reference [23]. A reduced-bit multiplication algorithm based on Urdhva Tiryagbhyam algorithm is described in [9]. Reference [9] gives the implementation of $4 \times 4$ bit multiplier, but this literature provides a better-optimized hardware architecture of Urdhva Tiryagb-hyam algorithm. Various efficient multiplication algorithms used in floating-point multipliers are described in papers [12,13,20]. A high-speed binary floating-point multiplier based on Dadda algorithm is presented in paper [13]. A Wallace tree multiplier using Booth recoder for fast arithmetic circuits on FPGA is proposed in [20], and it gives an improved version of tree-based Wallace tree multiplier architecture. Architecture for a fast 32-bit floating-point multiplier compliant with the single-precision IEEE 754-2008 standard is proposed in [12], and this design intends to make the multiplier faster by implementing adders having the least power delay constant, which helps in reducing the delay caused by the propagation of the carry.

Papers [3,18,19] deal with the binary multiplication based on Vedic mathematics. Paper [22] describes an efficient implementation of an IEEE 754 single-precision floating-point multiplier using Vedic mathematics. A VLSI implementation of low-power $16 \times 16$ bit multiplier using Vedic mathematics method is described in paper [3]. An excellent hardware implementation of Urdhva Tiryagbhyam algorithm is illustrated in this article. An $8 \times 8$ bit multiplier using Urdhva Tiryagbhyam algorithm is presented in paper [19]. It gives a better hardware implementation in terms of delay. Some features of some of these papers are adopted for the proposed model. Also, a comparison of those papers with the proposed Karatsuba–Urdhva algorithm is made regarding delay and area and observed that the proposed model is a better implementation.

## 3 Design of Proposed Model

The proposed model is a run-time-reconfigurable floating-point matrix multiplier, which can be used for high-speed and multiple precision applications. The basic unit of the model is the processing element (PE), which is a full-fledged $2 \times 2$ matrix multiplier. The design is made highly parallel with PE as the core processing unit. The model uses Strassen algorithm as the matrix multiplication algorithm. Strassen algorithm divides the operand matrices into sub-matrices of order $2 \times 2$, and the PE performs calculations on these sub-matrices. The model is made highly parallel such that all the sub-matrix multiplication operations for a $4 \times 4$ matrix will be executed in parallel, and hence, it will take only the execution time of a $2 \times 2$ matrix multiplication to multiply a $4 \times 4$ matrix. This parallel execution saves the execution time by more than eight times because a $4 \times 4$ matrix will have 64 multiplication and 48 addition operations, whereas a $2 \times 2$ matrix multiplication operation takes only eight multiplication and four addition operations. By using Strassen algorithm, the number of multiplications is reduced, which further increases the efficiency. The block diagram of the proposed model is shown in Fig. 1.

The proposed model has two levels. The top level is the Strassen algorithm, and the bottom level is the proposed run-time-reconfigurable multi-precision floating-point multiplier. Strassen algorithm is used for matrix operations, and run-time-reconfigurable multi-precision floating-point multiplier is used to multiply matrix elements inside the processing element. The main blocks of the proposed model are explained in detail in the following sections.

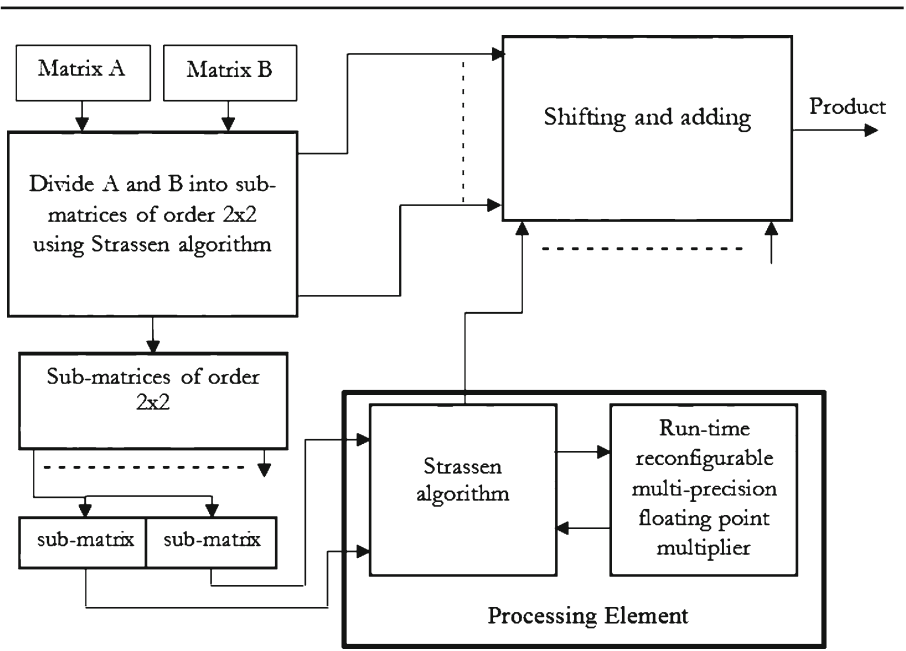

**Fig. 1** Block diagram of the proposed model

### 3.1 Strassen Algorithm

Strassen algorithm was introduced by Strassen in 1969, which required fewer multiplications of matrix elements than the classical matrix multiplication approach [6]. It is faster than the standard matrix multiplication algorithm, but slower than the fastest algorithm, Coppersmith–Winograd algorithm. Even though Coppersmith–Winograd algorithm and its derived algorithms are faster, it is not feasible to implement Coppersmith–Winograd algorithm practically. Coppersmith–Winograd algorithm is frequently used as a building block in other algorithms to prove theoretical time bounds, and hence, it is not used in practice. Coppersmith–Winograd algorithm only provides an advantage for matrices which are so large that they cannot be processed by modern hardware.

Strassen showed that a $2 \times 2$ matrix multiplication can be performed only with seven multiplications and 18 additions or subtractions. Strassen's algorithm is a divide-and-conquer algorithm which partitions each of the operand matrices into sub-matrices of equal size and employs a divide-and-conquer strategy. It divides any $n \times n$ matrix into sub-matrices, and each sub-matrix is of dimension $\frac{n}{2}$, where $n$ is the size of the matrix. It is the first algorithm to break the $n^3$ 'barrier'.

Let $A$, $B$ be the two matrices of size $4 \times 4$. Matrix $A$ is divided into sub-matrices $A_0$, $A_1$, $A_2$ and $A_3$ of size $2 \times 2$, and matrix $B$ is divided into sub-matrices $B_0$, $B_1$, $B_2$ and $B_3$ of size $2 \times 2$. Let the product matrix be $C$. Let $C_0$, $C_1$, $C_2$ and $C_3$ be the sub-matrices of matrix $C$, which are also of size $2 \times 2$. $C_0$, $C_1$, $C_2$ and $C_3$ can be determined from Eq. (1).

**Fig. 2** Strassen's algorithm illustration

| A | | x | B | | = | C | |
|---|---|---|---|---|---|---|---|
| A₀ | A₁ | X | B₀ | B₁ | = | A₀.B₀+A₁.B₂ | A₀.B₁+A₁.B₃ |
| A₂ | A₃ | | B₂ | B₃ | | A₂.B₀+A₃.B₂ | A₂.B₁+A₃.B₃ |

$$\left.\begin{aligned} C_0 &= A_0 \cdot B_0 + A_1 \cdot B_2 \\ C_1 &= A_0 \cdot B_1 + A_1 \cdot B_3 \\ C_2 &= A_2 \cdot B_0 + A_3 \cdot B_2 \\ C_3 &= A_2 \cdot B_1 + A_3 \cdot B_3 \end{aligned}\right\} \quad (1)$$

The process is illustrated in Fig. 2.

Let matrix $A_0 = \begin{pmatrix} a_{11} & a_{12} \\ a_{21} & a_{22} \end{pmatrix}$ and matrix $B_0 = \begin{pmatrix} b_{11} & b_{12} \\ b_{21} & b_{22} \end{pmatrix}$.

Then, the partial product matrix $P$ is obtained as,

$$P = A_0 \cdot B_0, \text{ where } P = \begin{pmatrix} p_{11} & p_{12} \\ p_{21} & p_{22} \end{pmatrix}$$

i.e. $\begin{pmatrix} p_{11} & p_{12} \\ p_{21} & p_{22} \end{pmatrix} = \begin{pmatrix} a_{11} & a_{12} \\ a_{21} & a_{22} \end{pmatrix} \cdot \begin{pmatrix} b_{11} & b_{12} \\ b_{21} & b_{22} \end{pmatrix}$

The partial products are obtained as shown in operation (2)

$$\left.\begin{aligned} S_1 &= (a_{11} + a_{22}) \cdot (b_{11} + b_{22}) \\ S_2 &= (a_{21} + a_{22}) \cdot b_{11} \\ S_3 &= a_{11} \cdot (b_{12} - b_{22}) \\ S_4 &= a_{22} \cdot (b_{21} - b_{11}) \\ S_5 &= (a_{11} + a_{12}) \cdot b_{22} \\ S_6 &= (a_{21} - a_{11}) \cdot (b_{11} + b_{12}) \\ S_7 &= (a_{12} - a_{22}) \cdot (b_{21} + b_{22}) \end{aligned}\right\} \quad (2)$$

The product matrix elements are obtained using Eq. (3) as follows.

$$\left.\begin{aligned} p_{11} &= S_1 + S_4 - S_5 + S_7 \\ p_{12} &= S_3 + S_5 \\ p_{21} &= S_2 + S_4 \\ p_{11} &= S_1 - S_2 + S_3 + S_6 \end{aligned}\right\} \quad (3)$$

From Eq. (2), it can be noticed that Strassen's algorithm requires only seven multiplication operations to compute the result for a second-order matrix, whereas the conventional algorithm requires eight multiplications. Hence, the number of multiplications '$M$' required for a $n$th-order matrix according to Strassen algorithm is

$$M(n) = 7M\left(\frac{n}{2}\right), \quad M(1) = 7 \quad \text{for } n \geq 2 \quad (4)$$

where $M\left(\frac{n}{2}\right)$ is the multiplication function of sub-matrices of order $\frac{n}{2} \times \frac{n}{2}$.

The complexity of Strassen algorithm can be written as

$$T(n) = 7T\left(\frac{n}{2}\right) + cn^2 \quad \text{for } n \geq 2 \tag{5}$$

Each sub-matrix is of size $\frac{n}{2} \times \frac{n}{2}$, and there are 7 multiplication operations. Adding the matrices together will take $cn^2$ steps for some fixed constant $c$ (because a matrix has $n^2$ entries).

By applying Master theorem [10], Eq. (5) works out to be

$$T(n) = \Theta\left(7^{\log_2 n}\right) = \Theta\left(n^{\log_2 7}\right) = O\left(n^{2.81}\right) \tag{6}$$

The complexity of Strassen algorithm is $O(n^{2.81})$, which is better than the classical algorithm.

In classical multiplication algorithm, it takes 8 multiplication operations to compute the product of two $2 \times 2$ matrices. The equations for product matrix elements are shown in Eq. (7).

$$\left.\begin{aligned}
p_{11} &= (a_{11} \cdot b_{11}) + (a_{12} \cdot b_{21}) \\
p_{12} &= (a_{11} \cdot b_{12}) + (a_{12} \cdot b_{22}) \\
p_{21} &= (a_{21} \cdot b_{11}) + (a_{22} \cdot b_{21}) \\
p_{22} &= (a_{21} \cdot b_{12}) + (a_{22} \cdot b_{22})
\end{aligned}\right\} \tag{7}$$

Hence, according to the classical algorithm, a $n$th-order matrix has a complexity $\Theta(n^{\log_2 8}) = O(n^3)$.

There are two options to be able to use bigger matrices: bottom-up and top-down.

*Bottom-up* In this method, the two matrices of size $n$ ($n = 2p$, where $p > 1$) are divided into sub-matrices of size $2^q$, $q < p$. A smaller matrix of size $m \times m$ ($m = 2^{p-q}$) is obtained. The multiplication of these two matrices—the external algorithm—is carried out using the classical algorithm, while the two matrices of size $2^q \times 2^q$ are multiplied by using the Strassen algorithm—internal algorithm.

*Top-down* This method changes the methods used in the bottom-up design. It uses Strassen as external algorithm and the classical multiplication as the internal algorithm. Bottom-up design is not working viable with an FPGA because the number of multiplications needed for the internal algorithm is too high. On the other hand, the top-down alternative is suitable. Besides, this method allows a pipelined hardware structure, improving the performance.

A variant of this top-down algorithm is shown in operation (8). It allows for starting the multiplication before all the coefficients of the matrix are already found out. For this to be possible, all the terms shown in (9) must be changed to $\alpha$ and $\beta$ as given in operation (9).

$$\left.\begin{aligned}
S_{ij}^1 &= \sum_{k=1}^{m} \alpha_{ik}^1 \beta_{kj}^1 \\
S_{ij}^2 &= \sum_{k=1}^{m} \alpha_{ik}^2 \cdot b_{2k-1,2j-1} \\
S_{ij}^3 &= \sum_{k=1}^{m} a_{2k-1,2j-1} \cdot \beta_{kj}^2 \\
S_{ij}^4 &= \sum_{k=1}^{m} a_{2k,2j} \cdot \beta_{kj}^3 \\
S_{ij}^5 &= \sum_{k=1}^{m} \alpha_{ik}^3 \cdot b_{2k,2j} \\
S_{ij}^6 &= \sum_{k=1}^{m} \alpha_{ik}^4 \cdot \beta_{kj}^4 \\
S_{ij}^7 &= \sum_{k=1}^{m} \alpha_{ik}^5 \cdot \beta_{kj}^5
\end{aligned}\right\} \quad (8)$$

$$\left.\begin{aligned}
\alpha_{ik}^1 &= a_{2i-1,2k-1} + a_{2i,2k} \\
\alpha_{ik}^2 &= a_{2i,2k-1} + a_{2i,2k} \\
\alpha_{ik}^3 &= a_{2i-1,2k-1} + a_{2i-1,2k} \\
\alpha_{ik}^4 &= a_{2i,2k-1} - a_{2i-1,2k-1} \\
\alpha_{ik}^5 &= a_{2i-1,2k} - a_{2i,2k} \\
\beta_{kj}^1 &= b_{2k-1,2j-1} + b_{2k,2j} \\
\beta_{kj}^2 &= b_{2k-1,2j} - b_{2k,2j} \\
\beta_{kj}^3 &= b_{2k,2j-1} - b_{2k-1,2j-1} \\
\beta_{kj}^4 &= b_{2k-1,2j-1} + b_{2k-1,2j} \\
\beta_{kj}^5 &= b_{2k,2j-1} - b_{2k,2j}
\end{aligned}\right\} \quad (9)$$

### 3.2 Processing Element (PE)

The proposed model uses multiple processing elements (PEs) in parallel, where each processing element is a full-fledged $2 \times 2$ matrix multiplier. Each PE has two levels. The top level is the matrix multiplication algorithm, and the bottom level is the run-time-reconfigurable multi-precision floating-point multiplier. The matrix multiplication algorithm used is Strassen algorithm as in the top level of the proposed model. Run-time-reconfigurable multi-precision floating-point multiplier at the bottom level multiplies the individual matrix elements. The block diagram of the PE is shown in Fig. 3. Matrix *A* and matrix *B* are the input matrices, and matrix *C* is the output product matrix. Various blocks in the PE are explained in the following sections.

#### 3.2.1 Input and Output Registers

The registers are used to store the input and output matrix elements. It has 12 registers of 64 bit: Eight registers are used to store the input and four registers are used to store the output of the PE. It also has a 3-bit input register to store the precision select input.

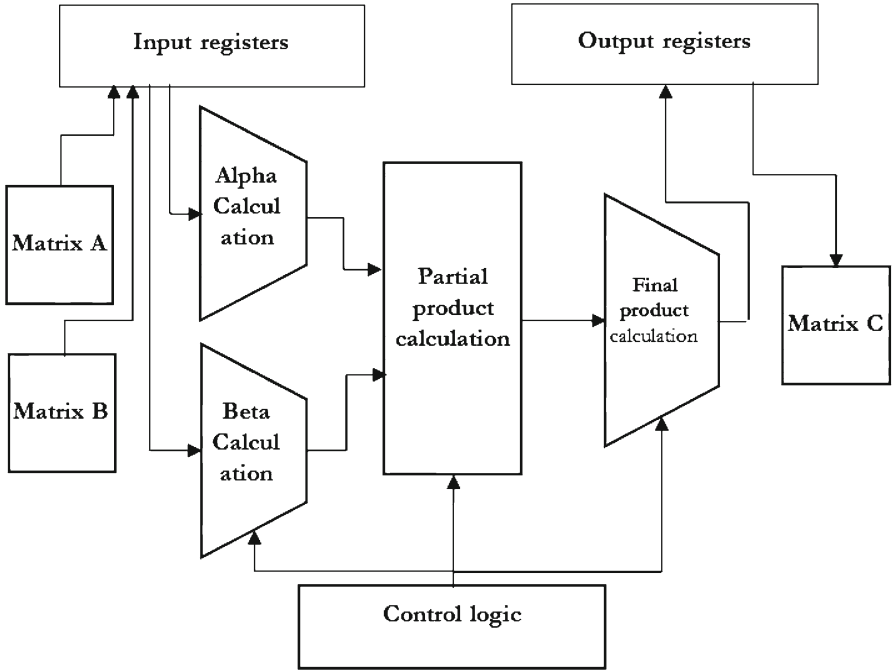

**Fig. 3** Block diagram of processing element

#### 3.2.2 Control Logic

The control logic is used to synchronize and control the operations. Two control signals, ready and reset, are used to control the execution of processes by the processing element. The control logic is effective mostly in the matrix element multiplication, i.e. in the run-time-reconfigurable multi-precision floating-point multiplier unit.

#### 3.2.3 Alpha and Beta Calculation

Alpha and beta calculation unit is the part of the implementation of Strassen algorithm. Alpha ($\alpha_{ik}^1$ to $\alpha_{ik}^5$) and beta ($\beta_{kj}^1$ to $\beta_{kj}^1$) values are calculated using the matrix elements, and these values are given to the partial product calculation unit to calculate partial products.

#### 3.2.4 Partial Product Calculation

This is also a unit corresponding to Strassen algorithm implementation. The corresponding alpha and beta values from alpha and beta calculation unit is used to calculate partial products $S_{ij}^1$ to $S_{ij}^7$. These partial products are calculated using the run-time-reconfigurable multi-precision floating-point multiplier according to the precision select input given to the PE. The partial product values are then provided to the product calculation unit.

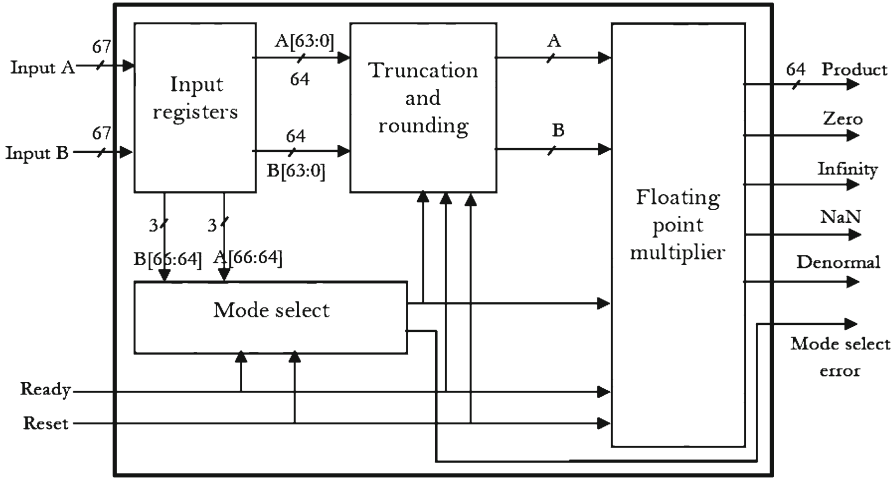

**Fig. 4** Block diagram of run-time-reconfigurable variable-precision floating-point multiplier

### 3.2.5 Product Calculation

The partial product values obtained from the previous block is used to calculate the complete result of the PE. The final result is calculated by the addition/subtraction of partial products according to the Strassen algorithm.

### 3.3 Run-Time-Reconfigurable Multi-Precision Floating-Point Multiplier

The proposed run-time-reconfigurable multi-precision floating-point multiplier performs the multiplication of matrix elements in the processing element. Multiplication of matrix elements in double-precision floating-point format is carried out according to the different precision requirements of the output. Since it performs variable-precision multiplication, it plays a significant role in the proposed model. The basic block of run-time-reconfigurable multi-precision floating-point multiplier is a double-precision floating-point unit. According to the precision required at the output, the size of the mantissa is varied in the floating-point multiplication operation. The block diagram of the model is shown in Fig. 4.

Run-time-reconfigurable multi-precision floating-point multiplier accepts two 67-bit operands and ready and reset signals as inputs. The 67-bit input is in a modified floating-point format where the first 3 bits are mode/precision select bits and the rest 64 bits are the operand to be multiplied in IEEE double-precision floating-point format. The precision select bits determine the precision and mantissa size of the floating-point multiplier to be chosen. According to the precision selected, the double-precision format input is truncated to custom mantissa size floating-point formats.

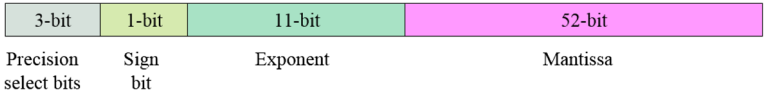

**Fig. 5** Modified floating-point format used in the model

Once the precision mode is selected, the input mantissa is truncated and rounded using the special rounding scheme used as explained in Sect. 3.3.4. This new rounding scheme ensures the least variation in results. The truncated input is given to the floating-point multiplier which uses multiple binary multipliers of different word lengths, from which a proper binary multiplier unit is selected based on the precision required and the rest of the multipliers are disabled and shut down. Only the required multiplier will be ON, and this ensures the least consumption of power. Once the multiplication is done, the result is made available at the output in double-precision floating-point format.

### 3.3.1 Modified Floating Point Format

For the purpose of run-time reconfiguration, the standard IEEE floating-point format is altered to fit the proposed model. The modified format we used is shown in Fig. 5. The multiplier accepts two inputs of 67 bits. The first 3 bits (66th bit to 64th bit) are the mode select bits. Mode select bits are used for selecting the appropriate precision mode for the application. The value of the mode select bits for both the inputs must be the same. Otherwise, a mode select error signal will be generated and the execution will be halted. The rest of the bit sizes are same as the IEEE-754 floating-point format. The 63rd bit is the sign bit, next 11 bits from 62nd bit to 52nd bit are the exponent bits, and the last 52 bits are mantissa bits.

In custom precision formats for different modes, the basic double-precision format is used with variable mantissa sizes. The different custom precision formats for different precision modes are shown in Fig. 6. Since mantissa multiplication is the only complex operation, it is done separately according to the mantissa sizes and all other operations are same as double-precision floating-point operations. All these custom precision operations are performed by the device itself, and hence, users cannot access or change the custom precision formats. Only the end result, which is of double-precision floating-point format, is outputted.

Even though it uses a modified floating-point format, it does not alter the basic double-precision floating-point format and floating-point operations. Also, the output will be of IEEE double-precision floating-point format. The only change which is done with the IEEE format is that the three precision select bits are appended as the most significant bits. These precision select bits can either be generated by an application program or can be taken as a preset value for a particular application. While giving the inputs for multiplication, the precision select bits are appended to the double-precision format by the application program. The application program can be any software program, for example the automatic resolution/bit-rate selection of internet video streaming based on the connection speed.

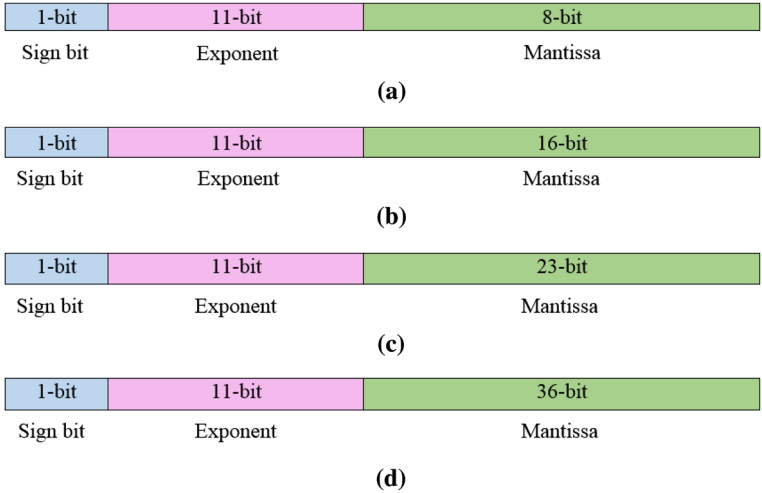

**Fig. 6** Custom precision floating-point formats used in the model. **a** Custom precision format 8-bit mantissa, **b** custom precision format 16-bit mantissa, **c** single-precision format, **d** custom precision format 32-bit mantissa

### 3.3.2 Inputs and Outputs

The run-time-reconfigurable multi-precision floating-point multiplier accepts four inputs and gives out six output signals. The inputs are the two operands, a ready and a reset signal. The the two operands are 67-bit wide. The outputs of the module are the product, mode select error signal, and four special values of the product, namely zero, infinity, not-a-number (NaN) and denormals. The product is of 64-bit-wide double-precision floating-point format, and all other output signals are of single bit.

### 3.3.3 Modes of Operation

The different mode select bit combinations for different modes are shown in Table 1. The various modes in the proposed multi-precision multiplier are the following. The modes are selected according to the mode select bits.

> *Mode 1*: Mode 1 is auto-mode, i.e. the controller itself will select the optimum mode by analysing the inputs and will start execution. The optimum mode is selected by counting the number of zeroes after a leading 1. If the number of zeroes is 6 or more after a leading 1, then the bits up to that leading 1 is counted. If the number of bits up to that leading 1 is <8, then mode 2 or 8-bit mantissa mode will be selected. If the number of bits before the leading 1 is <16, 16-bit mantissa mode will be selected and so on. The flow chart of auto-mode selection is shown in Fig. 7.

**Table 1** Mode select bits for different modes

| Mode | Mode selection bits | Precision (mantissa) |
| --- | --- | --- |
| Mode 1 (auto-mode) | 000 | According to input |
| Mode 2 | 001 | 8-bit |
| Mode 3 | 010 | 16-bit |
| Mode 4 | 011 | 23-bit |
| Mode 5 | 100 | 36-bit |
| Mode 6 | 101 | 52-bit |

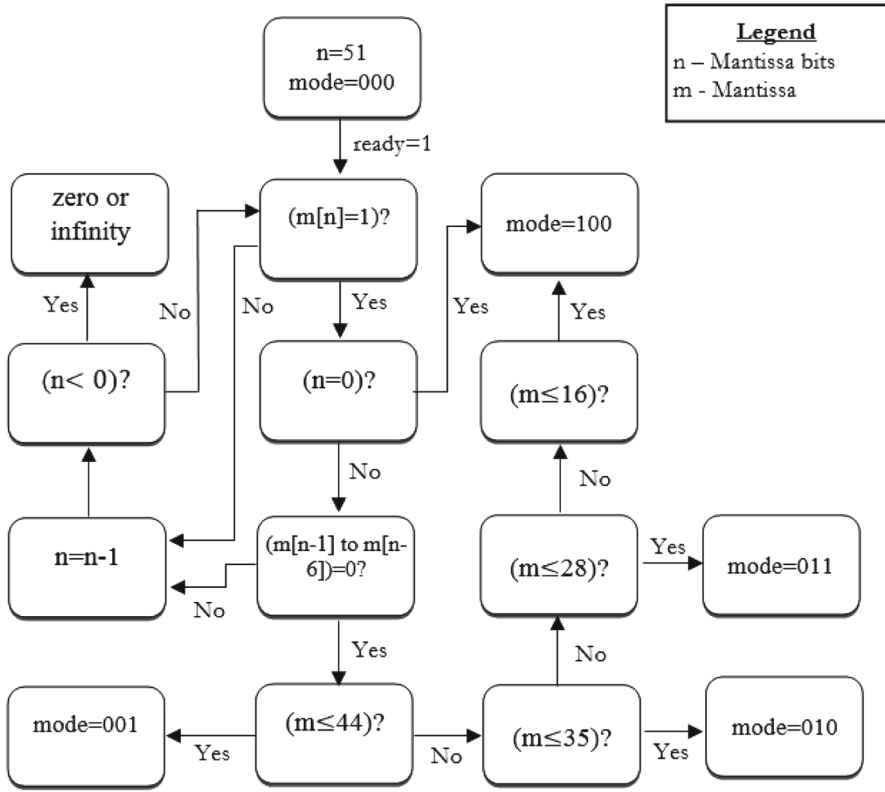

**Fig. 7** Flow chart of auto-mode selection

*Mode 2*: This is a custom precision format. It uses a basic double-precision floating-point multiplier with a mantissa size of 8 bits.
*Mode 3*: This is a custom precision format. It uses a basic double-precision floating-point multiplier with a mantissa size of 16 bits.
*Mode 4*: This is a custom precision format. It uses a basic double-precision floating-point multiplier with a mantissa size of 23 bits.

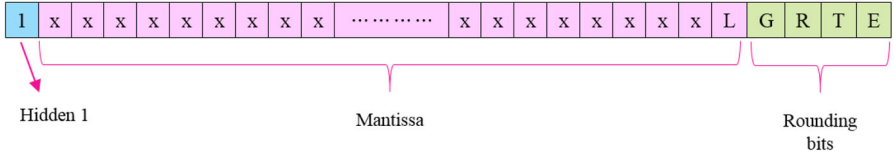

**Fig. 8** Rounding bits and mantissa of possible result

*Mode 5*: This is a custom precision format. It uses a basic double-precision floating-point multiplier with a mantissa size of 36 bits.
*Mode 6*: This mode is a full-fledged double-precision floating-point multiplier with accuracy at its best.

The modes with less number of mantissa bits are faster and consume less amount of power. These modes are best suited for integer multiplication and also for applications where accuracy is not a big issue. Rounding of bits is done before multiplication for every mode except for mode 6, and this reduces huge variations in results.

### 3.3.4 Truncation and Rounding

Truncation and rounding operations are done before and after multiplication except in mode 6. In mode 6, truncation and rounding are done only after multiplication. In all other custom precision modes, truncation of inputs is done according to the precision mode selected and rounding is done using the special rounding scheme developed. This special rounding scheme uses four bits instead of the conventional guard (G), round (R) and sticky bits (T). The additional bit used is termed as extra bit (E). The mantissa of a possible result with rounding bits is shown in Fig. 8.

The rounding scheme used is a round-up scheme, i.e. the bit 'rnd' is added to the LSB of the mantissa 'L'. The value of 'L' is calculated according to the least change in value determined when truncated. After careful evaluation of rounding bits, 'rnd' is calculated as in Eq. (10).

$$\text{rnd} = G\,\&\,(R\,|\,T\,|\,E) \tag{10}$$

This value or 'rnd' gives the least change in accuracy when rounded.

### 3.3.5 Floating-Point Multiplier

A floating-point number is represented in the IEEE-754 format [11] as $\pm s \times b^e$ or $\pm$significand $\times$ base$^{\text{exponent}}$. To perform multiplication of two floating-point numbers $\pm s1 \times b^{e1}$ and $\pm s2 \times b^{e2}$, the significand or mantissa of the numbers are multiplied to get the product mantissa, and exponents are added to get the product exponent, i.e. the product is $\pm(s1 \times s2) \times b^{(e1+e2)}$. The various conditions of special values and exceptions must be considered while multiplying two floating-point numbers. The hardware block diagram of floating-point multiplier is shown in Fig. 9.

In the proposed model, multiple precision is required and hence the floating-point multiplier used is different from the conventional floating-point multiplier and is shown in Fig. 10. It uses multipliers of different precisions according to the mode selected. The unused units are shut down to save power. The inputs A and B shown in Fig. 10 are of double-precision floating-point format and are truncated according to the mode selected before giving to the floating-point multiplier.

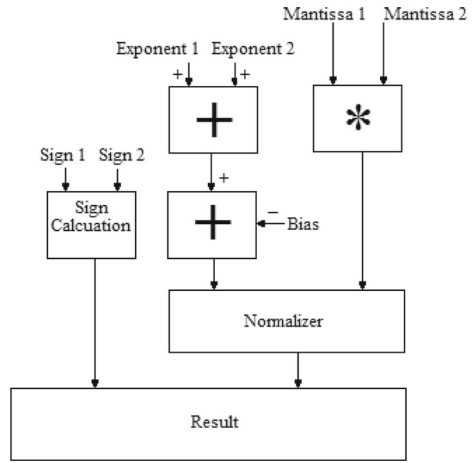

**Fig. 9** Floating-point multiplier block diagram

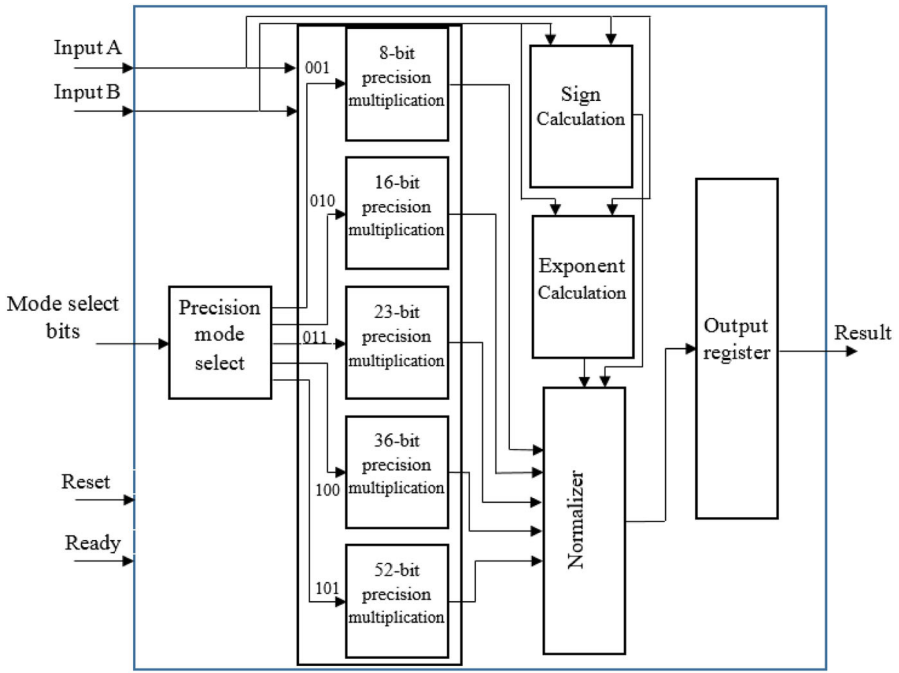

**Fig. 10** Block diagram of floating-point multiplier model in the proposed design

The important blocks in the implementation of proposed floating-point multiplier are described in detail in the following sections.

*3.3.5.1. Sign Calculation* The MSB of floating-point number represents the sign bit. The sign of the product will be positive if both the numbers are of the same sign and will be negative if numbers are of opposite sign. So, to obtain the sign of the product, a simple XOR gate can be used as the sign calculator.

*3.3.5.2. Addition of Exponents* The input exponents are added together to get the product exponent. Since a bias is used in the floating-point format exponent, it is required to subtract the bias from the sum of exponents to get the actual exponent. The value of bias is $127_{10}(01111111_2)$ for the single-precision format and $1023_{10}(01111111111_2)$ for the double-precision format. In the proposed custom precision format also, a bias of $1023_{10}$ is used. The computational time of mantissa multiplication operation is much more than the exponent addition. So a simple ripple carry adder and ripple borrow subtracter are optimal for exponent addition.

*3.3.5.3. Karatsuba–Urdhva Tiryagbhyam Binary Multiplier for Mantissa Multiplication* In floating-point multiplication, most important and complex part is the mantissa multiplication. Multiplication operation requires more time compared to addition operation. And as the number of bits increases, it consumes more area and time. In double-precision format and single-precision format, a $53 \times 53$ bit multiplier and a $24 \times 24$ bit multiplier are required, respectively, for mantissa multiplication. It requires much time to perform these operations, and it is the major contributor to the delay of the floating-point multiplier. The binary multiplier must be designed and implemented efficiently to reduce the area and delay constraints of the floating-point multiplier because binary multiplication is the most area and time-consuming operation as compared to addition. Since the proposed model uses multiple multipliers, the multiplication algorithm must be very much efficient.

To make the multiplication operation more area efficient and faster, a combination of Karatsuba algorithm [16] and Urdhva Tiryagbhyam algorithm [23] is used in the proposed model. A better and efficient implementation is done by combining the features of both these algorithms. The problem with conventional multiplication algorithm and Booth algorithm is that area increases drastically with increase in word length. Karatsuba algorithm and Urdhva Tiryagbhyam algorithm are good at their aspects, but have limitations too. Karatsuba algorithm uses a divide-and-conquer approach where it breaks down the inputs into most significant half and least significant half, and this process continues until the operands are 8-bit wide. Karatsuba algorithm is best suited for operands of higher bit length. But at lower bit lengths, it is not as efficient as it is at higher bit lengths.

Urdhva Tiryagbhyam algorithm is the best algorithm for binary multiplication regarding area and delay. But the partial products are added in a ripple manner in this algorithm, and hence as the number of bits increases, the delay also increases. For example, for a 4-bit multiplication, it requires six adders connected in a ripple manner. And an 8-bit multiplication requires 14 adders and so on. Compensating the delay will cause an increase in area. So Urdhva Tiryagbhyam algorithm is not that optimal if the number of bits is high.

To eliminate the limitations of Karatsuba algorithm, Urdhva Tiryagbhyam algorithm can be used at the lower stages where it is more efficient for multiplication of lower bit lengths. If Karatsuba algorithm is used at higher stages and Urdhva Tiryagbhyam algorithm is used at lower stages, it can somewhat compensate the limitations of both the algorithms and hence the multiplier becomes more efficient. The circuit is further optimized by using carry select and carry save adders instead of ripple carry adders. The usage of carry save adders reduces the delay to a great extent with minimal increase in hardware. These two algorithms are explained in detail in the following sections for better understanding. The model of Karatsuba–Urdhva Tiryagbhyam algorithm is shown in Fig. 11.

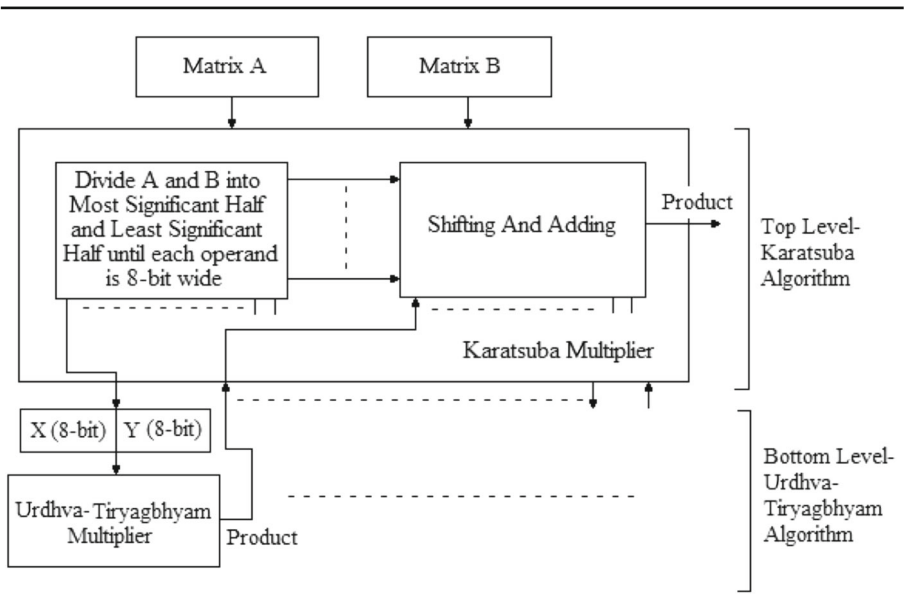

**Fig. 11** Karatsuba–Urdhva multiplier model

The two inputs A and B are two *n*-bit binary numbers. The inputs are divided into most significant half and least significant half until the two inputs are 8-bit wide. If 'n' is an even number and also after each division if the most and least significant bits are even, the algorithm uses Eq. (18). If the number of bits 'n' is odd or the division causes odd bit operands, the algorithm uses Eq. (20) for further steps. These 8-bit-wide most significant and least significant half operands are multiplied using Urdhva Tiryagbhyam algorithm. Finally, after getting all the sub-multiplication results, shifting of the results is done instead of multiplication by the powers of 2 as in Eqs. (18) or (20), and the results are added to get the final result.

There are efficient multiplication algorithms such as Booth and modified Booth algorithm, but the area requirement increases drastically when the number of operand bits increases. This proposed algorithm is very much efficient in terms of area when compared to other faster algorithms if the number of bits is high.

*Urdhva Tiryagbhyam Algorithm for Multiplication* Urdhva Tiryagbhyam sutra is an ancient Vedic mathematics method for multiplication. It is a general formula applicable to all cases of multiplication. The formula is very short and consists of only one compound word and means 'vertically and crosswise'. In Urdhva Tiryagbhyam algorithm, the number of steps required for multiplication can be reduced and hence the speed of multiplication can be increased. An illustration of steps for computing the product of two 4-bit numbers is shown in Eq. (11). The two inputs are $a_3a_2a_1a_0$ and $b_3b_2b_1b_0$, and let $p_7p_6p_5p_4p_3p_2p_1p_0$ be the product. The temporary partial products are $t_0, t_1, t_2, \ldots$ and $t_6$. The partial products are obtained from the steps illustrated below. The line notation of the steps is shown in Fig. 12.

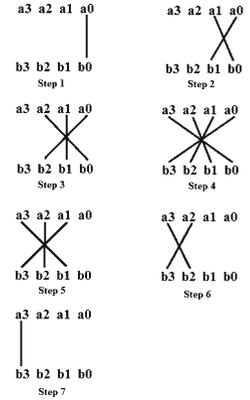

**Fig. 12** Line notation of Urdhva Tiryagbhyam sutra

$$\left.\begin{aligned}
&\text{Step1: } t_0 \text{ (1 bit)} = a_0b_0 \\
&\text{Step2: } t_1 \text{ (2 bit)} = a_1b_0 + a_0b_1 = t_1[1]\,t_1[0] \\
&\text{Step3: } t_2 \text{ (2 bit)} = a_2b_0 + a_1b_1 + a_0b_2 = t_2[1]\,t_2[0] \\
&\text{Step4: } t_3 \text{ (3 bit)} = a_3b_0 + a_2b_1 + a_1b_2 + a_0b_3 = t_3[2]\,t_3[1]\,t_3[0] \\
&\text{Step5: } t_4 \text{ (2 bit)} = a_3b_1 + a_2b_2 + a_1b_3 = t_4[1]\,t_4[0] \\
&\text{Step6: } t_5 \text{ (2 bit)} = a_3b_2 + a_2b_3 = t_5[1]\,t_5[0] \\
&\text{Step7: } t_6 \text{ (1 bit)} = a_3b_3
\end{aligned}\right\} \quad (11)$$

where $t_n[2], t_n[1], t_n[0]$ are the partial product bits according to binary positional weight.

The product is obtained by adding $s_1, s_2$ and $s_3$ as shown in Eq. (12), where $s_1, s_2$ and $s_3$ are the partial sum obtained.

$$\left.\begin{aligned}
s_1 &= t_6\,t_5[0]\,t_4[0]\,t_3[0]\,t_2[0]\,t_1[0]\,t_0 \\
s_2 &= t_5[1]\,t_4[1]\,t_3[1]\,t_2[1]\,t_1[1] \\
s_3 &= t_3[2]
\end{aligned}\right\} \quad (12)$$

$$\begin{array}{r}
\text{Product} = t_6 \; t_5[0] \; t_4[0] \; t_3[0] \; t_2[0] \; t_1[0] \; t_0 \\
+ \; t_5[1] \; t_4[1] \; t_3[1] \; t_2[1] \; t_1[1] \; 0 \\
+ \; t_3[2] \; 0 \quad\; 0 \quad\; 0 \quad\; 0 \quad\; 0 \\
\hline
p_7 \; p_6 \; p_5 \quad p_4 \quad\; p_3 \quad\; p_2 \quad\; p_1 \quad\; p_0
\end{array}$$

This method can be further optimized to reduce the number of hardware. A more optimized hardware architecture [9,19] is shown in Fig. 13. This model actually helps to eliminate the need for three-operand 7-bit adder and hence reduces hardware and delay. The adders are connected in a ripple manner.

The expressions for product bits are as shown in Eq. (13).

$$\left.\begin{aligned}
p_0 &= a_0 b_0 \\
p_1 &= \text{LSB of (Sum (ADDER1))} = \text{LSB of } (a_1 b_0 + a_0 b_1) \\
p_2 &= \text{LSB of (Sum (ADDER2))} = \text{LSB of (MSB(ADDER1)} + a_2 b_0 + a_1 b_1 + a_0 b_2) \\
p_3 &= \text{LSB of (Sum (ADDER3))} = \text{LSB of (MSB(ADDER2)} + a_3 b_0 + a_2 b_1 + a_1 b_2 + a_0 b_3) \\
p_4 &= \text{LSB of (Sum (ADDER4))} = \text{LSB of (MSB(ADDER1)} + a_3 b_1 + a_2 b_2 + a_1 b_3) \\
p_5 &= \text{LSB of (Sum (ADDER5))} = \text{LSB of (MSB(ADDER1)} + a_3 b_2 + a_2 b_3) \\
p_6 &= \text{LSB of (Sum (ADDER6))} = \text{LSB of (MSB(ADDER1)} + a_3 b_3) \\
p_7 &= \text{Carry of ADDER}
\end{aligned}\right\} \quad (13)$$

Since there are more than two operands in adders 2–5, we can use carry save addition to implement these adders. This technique reduces the delay to a great extent compared to the ripple carry adder.

*Karatsuba Algorithm for Multiplication* Karatsuba multiplication algorithm is best suited for multiplying very large numbers. This method was introduced by Anatoli Karatsuba in 1962. It is a divide-and-conquer method, in which we divide the numbers into their most significant half and least significant half and then multiplication is performed. Karatsuba algorithm reduces the number of multipliers required by replacing multiplication operations by addition operations. Additions operations are faster than multiplication operations, and hence the speed of the multiplier is increased. As the number of bits of inputs increases, Karatsuba algorithm becomes more efficient. This algorithm is optimal if the width of inputs is more than 16 bits. The hardware architecture of Karatsuba algorithm is shown in Fig. 14.

Karatsuba algorithm for two inputs $X$ and $Y$ can be explained as follows:
Product = $X \cdot Y$
$X$ and $Y$ can be written as,

$$X = 2^{n/2} \cdot X_l + X_r \quad (14)$$
$$Y = 2^{n/2} \cdot Y_l + Y_r \quad (15)$$

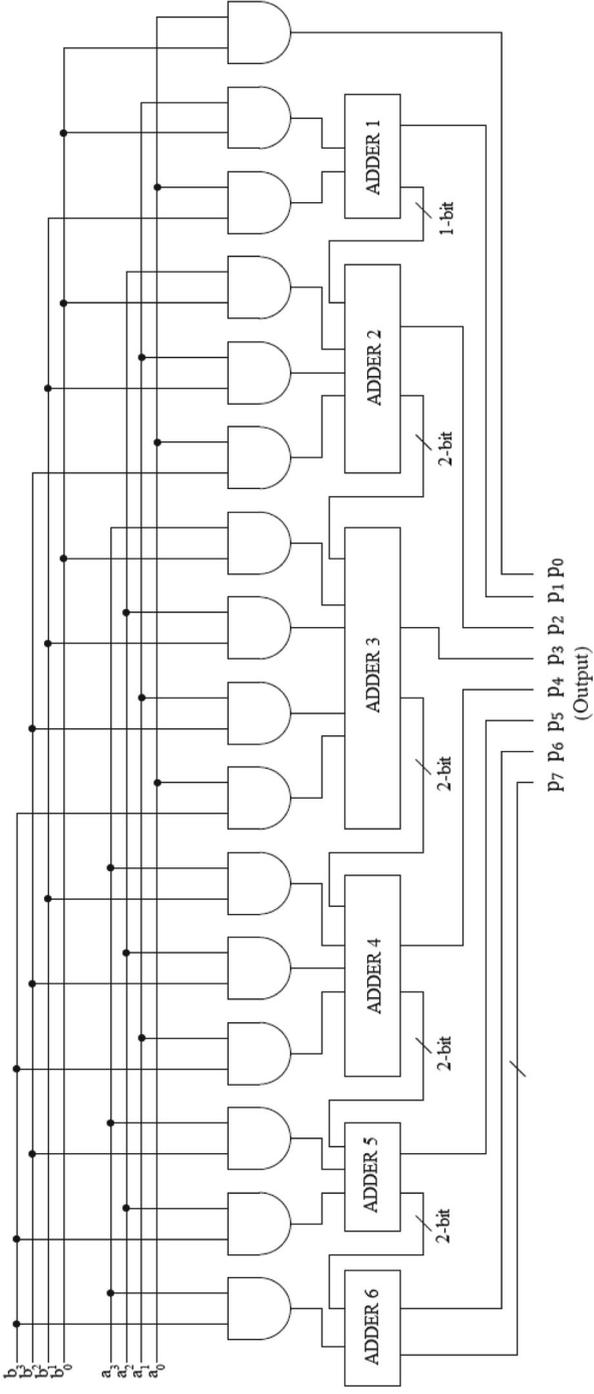

**Fig. 13** Hardware architecture of 4 × 4 Urdhva Tiryagbhyam multiplier

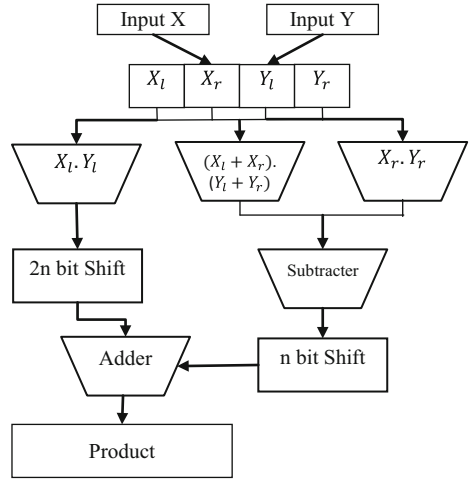

**Fig. 14** Hardware architecture of Karatsuba algorithm

where $X_l$, $Y_l$ and $X_r$, $Y_r$ are the more significant half and less significant half of $X$ and $Y$, respectively, and $n$ is the number of bits.

Then,

$$X \cdot Y = \left(2^{\frac{n}{2}} \cdot X_l + X_r\right) \cdot \left(2^{\frac{n}{2}} \cdot Y_l + Y_r\right)$$
$$= 2^n \cdot X_l Y_l + 2^{n/2}\left(X_l Y_r + X_r Y_l\right) + X_r Y_r \qquad (16)$$

The second term in Eq. (16) can be optimized to reduce the number of multiplication operations.

$$\text{i.e.;} \quad X_l Y_r + X_r Y_l = (X_l + X_r)(Y_l + Y_r) - X_l Y_l - X_r Y_r \qquad (17)$$

Equation (16) can be rewritten as,

$$X \cdot Y = 2^n \cdot X_l Y_l + X_r Y_r + 2^{\frac{n}{2}}\left((X_l + X_r)(Y_l + Y_r) - X_l Y_l - X_r Y_r\right) \qquad (18)$$

The recurrence of Karatsuba algorithm is as follows,

$$T(n) = 3T\left(\frac{n}{2}\right) + O(n) \hat{\approx} O\left(n^{1.585}\right) \qquad (19)$$

Even though these two algorithms are very straightforward to implement, some modifications are needed in the equations to make it easier to implement. The equations of Karatsuba algorithm work well for operands with even number of bits. When the operand bits are odd, division into more significant half and less significant half becomes difficult. The following set of equations gives the operation of operands with odd number of bits, which completely follows the actual method.

Let the number bits of two operands $X, Y$ be $n$, which is an odd number. Let us define two integers $f$ and $s$, where

$$f = \frac{n}{2}, \text{ integer part only.}$$
$$s = n - \frac{n}{2}$$

The length of $X_l$, $Y_l$ and $X_r$, $Y_r$ is $f$ and $s$, respectively.

Let $X_l Y_l = p_1$, $X_r Y_r = p_2$, $(X_l + X_r)(Y_l + Y_r) = p_3$. From Eq. (18), the result of multiplication can be written as

$$X \cdot Y = 2^{2s} \cdot (p_1) + p_2 + 2^s \cdot (p_3 - p_2 - p_1) \qquad (20)$$

*Normalization of the Result* Floating-point representations have a hidden bit in the mantissa, which always has a value 1, and hence it is not stored in the memory for saving one bit. A leading '1' in the mantissa is considered to be the hidden bit, i.e. the '1' just immediate to the left of decimal point. Usually, normalization is done by shifting so that the MSB of mantissa becomes nonzero, and nonzero means '1' in radix-2 representation. The decimal point in the mantissa multiplication result is shifted left if the leading '1' is not to the immediate left of decimal point. And for each left shift operation of the result, the exponent value is incremented by one. This is called normalization of the result. Since the value of the hidden bit is always 1, it is called 'hidden 1'.

*Representation of Exceptions* Some of the numbers cannot be represented with a normalized significand. To represent those numbers, a special code is assigned to it. The proposed model uses double-precision floating-point format, and we use four output signals, namely zero, infinity, not-a-number (NaN) and denormals to represent these exceptions. If the product has exponent + bias = 0 and significand = 0, then the result is taken as zero ($\pm 0$). If the product has exponent + bias = 1023 and significand = 0, then the result is taken as infinity ($\infty$). If the product has exponent + bias = 1023 and significand $\neq 0$, then the result is taken as NaN. Denormalized values or denormals are numbers without a hidden 1 and with the smallest possible exponent. Denormals are used to represent certain small numbers that cannot be represented as normalized numbers. If the product has exponent + bias = 0 and significand $\neq 0$, then the result is represented as denormal. Denormal is represented as $\pm 0.s \times 2^{-511}$, where $s$ is the significand or mantissa.

## 4 Implementation and Results

This project is programmed in Verilog HDL, synthesized and simulated using Xilinx synthesis tools (Xilinx ISE 14.7), targeted on Xilinx Virtex 5 ML10 Evaluation platform. We use a bottom-up design process to implement this project. The design is done in three stages. The first stage is the design and testing of Karatsuba–Urdhva

**Table 2** Performance analysis of various-word-length Karatsuba–Urdhva Tiryagbhyam binary multipliers

|  | 8-bit multiplier | 16-bit multiplier | 24-bit multiplier | 32-bit multiplier | 53-bit multiplier |
|---|---|---|---|---|---|
| Slices | 135 | 442 | 1259 | 1381 | 4587 |
| LUTs | 89 | 337 | 1013 | 1138 | 3891 |
| IOBs | 33 | 65 | 97 | 129 | 213 |
| Delay (ns) | 5.794 | 7.333 | 8.411 | 9.027 | 10.213 |
| $f_{max}$ (MHz) | 352.678 | 330.940 | 309.449 | 292.950 | 255.213 |

**Table 3** Comparison of 8-bit multipliers with the proposed multiplier

|  | Reference [18] | Reference [19] | Reference [22] | Proposed multiplier |
|---|---|---|---|---|
| Width | 8-bits | 8-bits | 8-bits | 8-bits |
| Delay (ns) | 28.27 | 15.050 | 23.973 | 9.396 |

**Table 4** Comparison of 16-bit multipliers with the proposed multiplier

|  | Reference [20]—Vedic multiplier | Reference [3] | Proposed multiplier |
|---|---|---|---|
| Width | 16-bits |  | 16-bits | 16-bits |
| Delay (ns) | 13.452 | 27.148 | 11.514 |

**Table 5** Comparison of 24-bit multipliers with the proposed multiplier

|  | Slices | LUTs | Delay (ns) |
|---|---|---|---|
| Reference [12] | 1306 | 2329 | 16.316 |
| Proposed multiplier | 972 | 1018 | 12.996 |

Tiryagbhyam binary multiplier. Binary multipliers for all the stages are designed and tested at this stage. The results of implementation are shown in tables.

Table 2 shows the performance analysis of various-bit-length Karatsuba–Urdhva Tiryagbhyam binary multipliers. Tables 3, 4, 5 and 6 show the comparison of some existing models of different word lengths with the proposed multiplier. Note that the comparison of area is done according to the number of LUTs used and the number of LUTs used actually vary with different coding styles used. Figure 15 shows the percentage change in area and delay with increasing bit length. The values are calculated by determining the relative change in area and delay when the size of the multiplier changes from 8 to 16 bits, 16 to 32 bits and so on. For example, when the size of the multiplier changes from 16 to 32 bits, the change in area is 3.3768 times the area of 16-bit multiplier. From this graph, it can be seen that the increase in percentage area is not significant and also the percentage increase in delay decreases with increasing bit length.

**Table 6** Comparison of 32-bit multipliers with the proposed multiplier

|  | LUTs | Delay (ns) |
| --- | --- | --- |
| Reference [20]—modified Booth multiplier (Radix-8) | 2721 | 12.081 |
| Reference [20]—modified Booth multiplier (Radix-16) | 7161 | 11.564 |
| Reference [20] | 2704 | 9.536 |
| Proposed multiplier | 1545 | 13.141 |

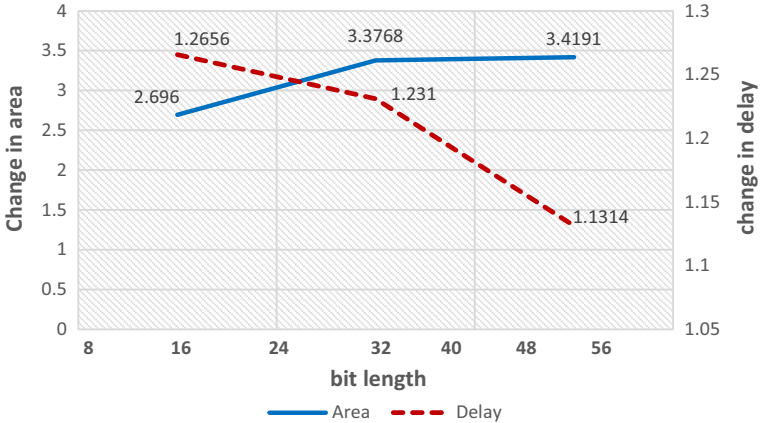

**Fig. 15** Percentage change in area and delay with change in word length of Karatsuba–Urdhva Tiryagbhyam multiplier

The second stage is the design and testing of individual floating-point units with different precision formats and integration of all the individual units to implement the run-time-reconfigurable multi-precision floating-point multiplier. The results and analysis of the implementation are discussed below.

Table 7 shows the performance analysis of different floating-point units having different precisions. It can be seen that with increasing word length, the increase in delay is significantly low. Table 8 shows the delay and area comparison of single-precision floating-point multiplier with single-precision FP multiplier in the proposed model. Even though the area is comparable to other models, the delay is much better in the proposed single-precision floating-point unit. Figure 16 shows the percentage variation in area and delay with increasing precision of floating-point units. The values are calculated by determining the relative change in area and delay when the size of the floating-point multiplier of different precisions changes. For example, when the precision changes from 8 to 16 bit and from 16 to 32 bit, the area changes 2.05 times and 2.39 times, respectively. It can be seen that the increase in area and delay is significantly less with the increase in word length.

The analysis results of multiplication of two double-precision floating-point numbers in different modes are shown in Table 9. From the table, it can be seen that the precision of mantissa of final result varies in lower precision modes, but only by around $\frac{1}{1000}$, which is good enough for integer-level precision and for small numbers with small exponents. The variation in precision of different modes is shown in Fig. 17. From the graph, it can be seen that the precision varies slightly in 8- and 16-bit modes but not much variation is there in other higher-order precision modes. So it can be concluded that the lower precision modes can be used for integer-level precision and high-level precision modes can be used for greater precisions.

**Table 7** Performance analysis of various floating-point units in the proposed run-time-reconfigurable multi-precision floating-point multiplier

|  | 8-bit precision floating-point multiplier | 16-bit precision floating-point multiplier | 23-bit precision floating-point multiplier | Double-precision floating-point multiplier |
|---|---|---|---|---|
| Slices | 157 | 497 | 1259 | 4587 |
| LUTs | 220 | 451 | 1078 | 3983 |
| IOBs | 61 | 83 | 104 | 193 |
| Delay (ns) | 7.234 | 10.234 | 11.008 | 12.785 |
| $f_{max}$ (MHz) | 344.767 | 323.630 | 309.449 | 255.213 |

**Table 8** Delay and area comparison of single-precision floating-point multiplier with single-precision FP multiplier in the proposed model

|  | Slices | LUTs | Delay (ns) |
|---|---|---|---|
| Reference [12] | 1269 | 2270 | 18.783 |
| Reference [13] | 1149 | 1146 | – |
| Proposed multiplier | 1259 | 1078 | 11.008 |

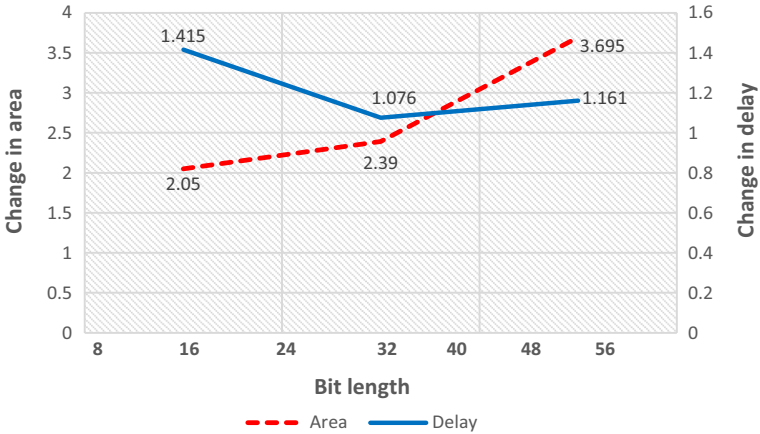

**Fig. 16** Percentage change in area and delay with change in word length of various floating-point units in the proposed run-time-reconfigurable multi-precision floating-point multiplier

**Table 9** Analysis of results of double-precision floating-point numbers in different modes

|  | Result in double-precision format | Value in decimal | Variation of mantissa in result |
|---|---|---|---|
| Input 1 | 4069b130ae804118 | $1.605759317 \times 2^7$ | – |
| Input 2 | 4069b130ae804118 | $1.605759317 \times 2^7$ | – |
| Auto-mode | 40e4a0b1337cdfbd | $1.289231492 \times 2^{15}$ | 0.0 |
| 8-bit precision | 40e49ec800000000 | $1.288978577 \times 2^{15}$ | 0.000252915 |
| 16-bit precision | 40e4a0b01b480000 | $1.289072997 \times 2^{15}$ | 0.000158495 |
| 23-bit precision | 40e4a0b11c33e320 | $1.289231405 \times 2^{15}$ | 0.000000087 |
| Double precision | 40e4a0b1337cdfbd | $1.289231492 \times 2^{15}$ | 0.0 |

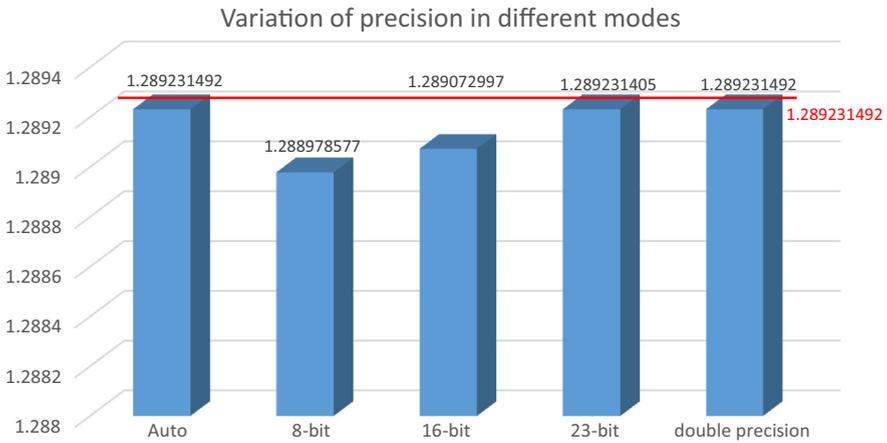

**Fig. 17** Variation of precision of multiplication results in different modes

Figure 18 shows the reduction in area while using run-time-reconfigurable multi-precision floating-point multiplier when compared to conventional double-precision floating-point multiplier. It can be seen that area usage for lower precision modes are very much less and this will drastically reduce the power consumption at lower precision modes. So if there are no constraints of precision, the device can be operated in lower precision modes with literally 'zero' power consumption.

The last stage in the design process is the implementation and testing of the complete run-time-reconfigurable multi-precision floating-point matrix multiplier. The complete implementation is done by implementing Strassen algorithm and integrating proposed run-time-reconfigurable multi-precision floating-point multiplier in the design. The top module of the processing element is shown in Fig. 19.

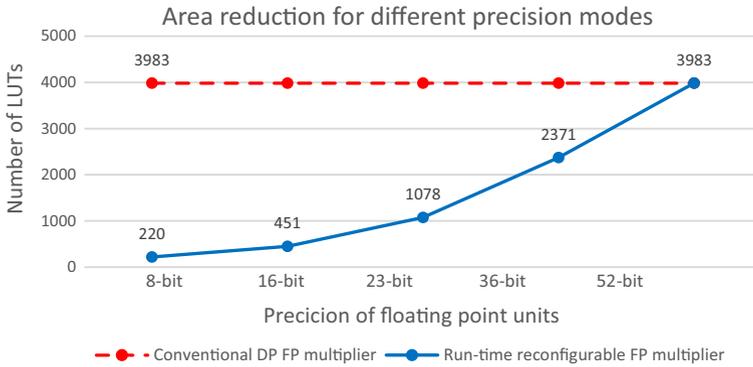

**Fig. 18** Reduction in area while using run-time-reconfigurable multi-precision floating-point multiplier when compared to conventional double-precision floating-point multiplier

**Fig. 19** Top module of processing element

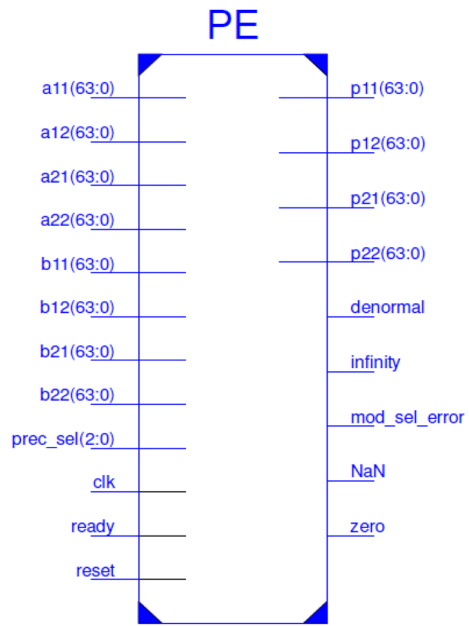

## 5 Conclusion and Future Scope

The result obtained from this work emphasizes the importance of run-time-reconfigurability and variable-precision requirements in electronic design. The results show that the delay and power consumption of the chip can be adequately adjusted according to different precision requirements. Also, the proposed design can reduce the percentage increase in area and delay with increasing word length of floating-point multipliers by using an efficient combination of Karatsuba algorithm and Urdhva Tiryagbhyam algorithm for implementing the unsigned binary multiplier. Even though it is efficient in terms of area, power and delay in the run-time environment, the overall area requirement of the model is more when compared to conventional designs. This is because of parallelism applied to the model. Hence, much improvement is needed to reduce the overall area requirement of the project. This design has broad applications in the area of image and signal processing where it can be used in portable devices and for military applications where saving battery power is very much important.